\documentclass[12pt,latex2e]{article}
\usepackage{enumerate}
\usepackage{pgf,pgfarrows,pgfnodes,pgfautomata,pgfheaps,pgfshade}
\usepackage{amsmath}
\usepackage{amsfonts}
\usepackage{bbm}
\usepackage{graphicx}
\usepackage{lmodern}
\newcommand{\rf}[1]{(\ref{eq:#1})}

\title{Cosmology of  Spin-2 Fields}
\author{
M. D. Maia\\
University of  Brasilia, Institute of Physics\\70910-970, Brasilia DF\\
maia@unb.br}
\begin{document}
\oddsidemargin -0.5cm
\evensidemargin -0.5cm
\textwidth 22cm
\topmargin -1.3cm
\textheight 24cm
\maketitle

\abstract{
The  Cauchy-Kowalevski theorem is  applied  to  the solutions of Einstein's  equations and to  cosmology. Three fundamental requirements of the theorem: the use of  analytic series;  the existence of  the boundary  surfaces;  and  the   setting of the independent initial data are revised,  using  methods of  geometric  analysis. It is  shown that during its  relativistic phase, the  standard model of the  universe is governed by   Einstein's  gravitation  described as  a  massless  spin-2  field,  but  it  is necessarily complemented  by  massive   spin-2 field, which responds for the dark sector of the universe.  On the other hand, at the inflationary phase, the exponential growth of the  volume of the universe  is shown to be consistent with a  thermal, non-relativistic expansion.  These two phases  are separated by  the  last inflationary  surface.

}

\section{Gravitation and   Spin-2 Fields}

The standard model of  the universe describes  a  causal sequence of space-like 3-dimensional surfaces of  the space-time manifold (for  short, these will be referred to simply as  surfaces) evolving along a global  time-like  line.  Current  observations  tell that   these surfaces are  flat in the Riemann sense.
 Furthermore,  the observations  of  supernovae type  Ia  show  that these surfaces are subjected to a non-vanishing tension,  resulting from the unbalance between ordinary gravitating matter and  some yet  unknown  dark force, producing  an acceleration in the universe expansion.
The purpose of  this  note is  to   show that these  observations  are consistent with the proper formulation of the Cauchy problem in  General Relativity.

The spin-statistics theorem of relativistic  quantum mechanics show that quantum systems  with integer spin obey the Einstein-Bose statistics and those with half integer  spins  obey the Fermi-Dirac  statistics. One of its  consequences  is  that Einstein's   gravitation  fits  the  Einstein-Bose  statistics  as  a  massless spin-2 field.  The  relation between the spin   and the classical degrees of freedom  of a  field  $D_f$  measured in the 3-dimensional space-like hypersurfaces of the space-time is    $D_f =
(2s +1)$ .
 Thus,   scalar,  vector   and  symmetric  tensor field   are  related  to  spin-$0$, spin-$1$ and  spin-$2$ quantum  fields respectively \cite{Fierz,Pauli}.

In compliance with the spin-statistics theorem, the Lagrangian for  a generic spin-2 field  defined by  a  symmetric, non-singular tensor  $H_{\mu\nu}$ was  described by   M. Fierz  and  W. Pauli   in the Minkowski space-time as
\begin{equation}
{\cal L}\! =\!\frac{1}{4} [
 H_{,\mu}H^{,\mu}\! - \!
H_{\nu\rho,\mu}H^{\nu\rho,\mu}\!  -\!2H_{\mu\nu}{}^{,\mu}H^{,\nu}\! +\!  2 H_{\nu\rho,\mu}H^{\mu\rho,\nu}
- m^2(H_{\mu\nu}H^{\mu\nu} -H^2)], \label{eq:FierzPauli}
\end{equation}
where $H=\eta^{\mu\nu}H_{\mu\nu}$  and
$m$ is the rest  mass of   the  field \cite{FierzPauli}.  Notice that in a  3-dimensional surface $H_{\mu\nu}$  would have six degrees of  freedom,  but one of them is eliminated  on the account that the state of smallest energy of the field occurs  when\footnote{Our convention: Greek indices  run from 1 to 4 and Latin indices run from 1 to 3.}  $ \eta^{\mu\nu}H_{\mu\nu} = H_i{}^i +\eta^{44}H_{44}= 0$.

The relation between  spin-2 fields and gravitation was first noted by S. N. Gupta
in 1954, observing that in the case $m = 0$,  the  Euler-Lagrange equations  derived from \rf{FierzPauli}  coincide with Einstein's gravitational wave equations  obtained from    the linear perturbation  $g_{\mu\nu} = \eta_{\mu\nu} + \gamma_{\mu\nu}$ of the Minkowski metric $\eta_{\mu\nu}$  (using the   de Donder coordinate gauge)
\begin{equation}
	\Box^2 \psi_{\mu\nu} =0,\;\;\;\; \psi_{\mu\nu}=\gamma_{\mu\nu}-\frac{1}{2}\gamma\eta_{\mu\nu}. \label{eq:masslesswave}
\end{equation}
In the sequence, Gupta reversed the linear approximation procedure, by adding successive perturbation  terms  to the Minkowski metric
\begin{equation}
   g_{\mu\nu}= \eta_{\mu\nu} + \gamma_{\mu\nu} +\gamma^2_{\mu\nu}...,
\end{equation}
while  at the same  time calculating the respective curvature tensors. Comparing the result with the corresponding  perturbations of the  energy-momentum tensor of  the source term, at the end Gupta recovered the full non-linear Einstein's equations
\begin{equation}
R_{\mu\nu}-\frac{1}{2}R g_{\mu\nu}=8\pi G T_{\mu\nu} \label{eq:EE},
\end{equation}
Here $g_{\mu\nu}$ is  regarded as  a  field  on the  Minkowski's space-time \cite{Gupta}. Thus, the tensor $R_{\mu\nu}$  and  the scalar  $R$ in those  equations  are formally defined as  the Ricci tensor  and  Ricci  scalar  respectively  calculated  from $g_{\mu\nu}$ as in Riemannian geometry. Therefore,  equations  \rf{EE}  describe  a non-linear spin-2, massless  field in Minkowski's  space-time,  which is completely equivalent to the  Euler-Lagrange equations derived  from the  Einstein-Hilbert action  ${\cal L} = R\sqrt{-g}$ in General Relativity,
  up to an additional  divergence-free  term.
In order to obtain the full theory of  General Relativity  it becomes  also necessary to identify  the gravitational potential $g_{\mu\nu}$ with the space-time metric  geometry,  and to  postulate   the principles of general covariance and  of  equivalence, which are compatible with Einstein's equations.

 The above   correspondence is not a  coincidence. As  it happens with the other fundamental interactions, it  provides the missing  field theoretic support  to  Einstein's  dynamical  equations, which  were otherwise  derived  by  pure  geometrical considerations \cite{Feynman,Fronsdal,Sivaran}.

An  immediate  consequence of the   spin-2  characteristics of gravitation is  that  any  alternative relativistic  gravitational  theory   where the  gravitational potential is  defined  by the space-time  metric  will be  necessarily  defined  by  the Einstein-Hilbert  action. For example, in the so called $F(R)$  alternative theories of gravitation, where $F$ is a  function of the   Ricci scalar $R$ derived from the  metric tensor  taken as  the gravitational field,  will necessarily lead to   $F(R)=R \pm (\mbox{a divergence-free  term})$.
To be sure,  it is  necessary to verify  that any  other spin-2 field described by  \rf{FierzPauli} does not lead to  a valid  alternative  theory of gravitation.  This was checked  in the 70's \cite{VanDam,Zakharov},
	by  repeating the same  non-linear reconstruction  introduced by Gupta,  but using  the Fierz-Pauli  Lagrangian \rf{FierzPauli} with $m\neq 0$ .
It  was soon found  that such theory meets  strong disagreements with the  classic gravitational experiments, resulting from the non-linear  interference of  the mass term in the resulting  equations. In  addition,  as  a  field theory in Minkowski's  space-time, it contains  tachionic and acausal solutions \cite{BoulwareDeser,DeserHenneaux,Deser2013,Kurt,CdeRham}.

It should  be added that  the  proposed  ``massive gravity''   cannot  have   Einstein's  gravity as  its  zero mass limit. This follows from the  fact that  in any Poincar\'e invariant field theory, the concept of  mass   is given  by the eigenvalues of the Casimir  mass operator, whose  spectrum is  composed of a distribution of discrete  isolated  points  in the real line \cite{Flato}. Such  spectrum  does not  contain an infinite number of  arbitrarily small masses, required to  define the zero mass limit. Consequently,  the  massless, non-linear spin-2 field (or particle) in Minkowski space defined by  Fierz-Pauli  is  isolated and independent  from the massive  Fierz-Pauli spin-2 field.  In the  following we  give a geometrical  interpretation for  this massive field.
	
\section{The Cauchy Problem in General Relativity}

The global hyperbolicity  of Einstein's  equations  implies that   its  causal  solutions can be  uniquely expressed by a  Cauchy sequence of   surfaces,  orthogonal  to an  integrable time-like  vector field,  satisfying the conditions of the Cauchy-Kowalevski theorem.
By an integrable vector field it is meant  that  through the points of each surface  passes  a curve whose velocity vector  is the said  vector field. The standard model of  cosmology  provides   a prime  example of the solutions of Einstein's  equations described by such  Cauchy sequence, composed by flat  space-like  surfaces  evolving in a  cosmic time.

 However, the conditions  established by  the Cauchy-Kowalevski theorem can be  quite  restrictive  for  high energy physics  and to General Relativity in particular. The  three most  important are:

(a) Since  all surfaces of a   space-time are submanifolds,  their  metric geometry are necessarily   induced by the space-time-metric \cite{Eisenhart}. Therefore  the geometry of the  Cauchy surfaces, can be defined only after the problem  is solved.

(b) In  General Relativity   all gravitational properties are derived from the metric, assumed to be   the only independent variable.  Consequently, the  velocity of propagation of  the surfaces, which is  to be set  as  one of  the  initial values,  is  also  dependent of the metric. For  example, using coordinates co-moving  with these surfaces,  the  velocity of propagation of the  surfaces   is   expressed   by the  metric affine connection as
\begin{equation}
\dot{g}_{ij}=  -2 g_{4\mu}\Gamma_{ij}^\mu \label{eq:Christoffelflow},
\end{equation}
 which is entirely dependent of the metric and its   use to  build the  Cauchy problem,  can lead  erroneous conclusions.

(c) Finally,  the standard proof of the Cauchy-Kowalevski theorem holds  for  analytic  functions of the coordinates. In General Relativity  and  in most non-linear differential  equations of physics the best  we may hope  is to use  differentiable  functions.  This problem is  actually  a  major  issue   in  the analysis of  Riemannian manifolds,   which has been  solved only recently with the  development of  ``Geometric Analysis'':  The   solutions of  non-linear differential equations by use of ``smooth deformations of  surfaces''.  Due to  the relevance  of  this concept  to  General Relativity, we  end  this  section with a  brief  introduction,  while referring  the  details   to  the   original cited  sources.

We find  it  more intuitive  to  start with
the  proof of the  Poincar\'e  Conjecture,  which shows how to  continuously  deform   a closed  compact surface into a sphere. The proof  uses thermodynamics  to  determine  the variation of the  volume enclosed by a compact surface with the  temperature. Assuming that  the  volume vary  exponentially with the temperature (or with the  entropy) $u$, $ \int \sqrt{g} dv =\int e^u dv $.
Replacing  	$\sqrt{g}= e^u$ in  Fourier's heat equation,  it follows that
$\frac{\partial u}{\partial t}=  \frac{1}{2}g^{\mu\nu}\frac{\partial g_{\mu\nu}}{\partial t}$.
Comparing this  result  with the expression of the Ricci tensor, written in geodesic coordinates,  after canceling the trace, we find the Ricci flow expression
\begin{equation}
\dot{g}_{ij} =-2 R_{\mu\nu},
\label{eq:Ricciflow}
\end{equation}
which  specifies  the velocity of the  ``deformation of  the geometry''   by the  Ricci tensor\cite{RHamilton}.
In 2002,  G.  Perelmann applied  \rf{Ricciflow}  to prove the Poincar\'e conjecture \cite{Perelmann}. Using   an intuitive approach, this can be  seen  in the following  way: Consider  a  sphere $\Sigma$ immersed  in a  3-dimensional  compact surface  $\bar{\Sigma}$, near  a  heat source. Next, mark the positions of the heat flow lines and smoothly push the surface $\bar{\Sigma}$,  so that  it always remains orthogonal to the heat flow lines, without creating wrinkles,  until it coincides with the enclosed sphere.

 		It can be  easily seen  that Perelmann's  result  cannot be   applied  to   General Relativity. Indeed, 		 by replacing \rf{Ricciflow} in Einstein's  equations it implies in a linear gravitation and in the vacuum,  gravitation would not propagate  at all.
However,  a similar  but more general concept of smooth  deformation of  surfaces was   developed in 1956 by  John Nash, applicable to  any  Riemannian manifold. Nash's original motivation  was  to  prove the existence  of  submanifolds  of  a  Riemannian  manifold  by  use of  differentiable  functions only \cite{Nash}.  Before Nash, the only known solutions of the  Gauss-Codazzi  equations,  which determines  the existence of  surfaces  in a  manifold,   requires the use of  analytic  series.
Nash's theorem  solved  this problem  by  applying a new concept of  smoothing  operators,  similar but  more general  than \rf{Christoffelflow} or  \rf{Ricciflow},  using  the extrinsic curvature of the  surface.  The extrinsic  curvature  gives a local  measure of  how  a surface  deviates locally from its tangent plane.  It is  therefore
a measure of  the local curvature
of  a surface  which is not entirely  dependent of the  surface metric.

The Nash  flow  expression can be  derived  from  the  propagation   of an  object $\bar{\Omega}$ defined in one surface $\bar{\Sigma}$, when it is Lie transported  to another surface.  That is,  along the integral curve of  the  orthogonal vector  field $\eta$:
$\Omega = \bar{\Omega} +t(\pounds_\eta \Omega)$,  where  $t$  is the parameter of the curve.
 In particular,  taking  $\bar{\Omega}$ to be  the metric $\bar{g}_{ij}$ and $\bar{k}_{ij}$ of the initial  surface,  we obtain in the new  surface  the  polynomial expansions \cite{Maia1985,Pirani}
\[
\begin{array}{ll}
g_{ij}(x, t) = \bar{g}_{ij} - 2t \bar{k}_{ij} + t^2 \bar{g}^{mn}\bar{k}_{im}\bar{k}_{jn},\vspace{2mm}\\
k_{ij}(x,t)  =
\bar{k}_{ij}\!\!-t\bar{g}^{mn}\bar{k}_{im}\bar{k}_{jn}.
\end{array}
\]
Comparing  the  derivative of the first expression with the second expression we easily obtain the  Nash Flow\footnote{In the ADM  formulation of  General Relativity  the  same  condition \rf{Nashflow} is  known as the York relation \cite{Bruhat}.}
\begin{equation}
\dot{g}_{ij}  =-2k_{ij}.  \label{eq:Nashflow}
\end{equation}

\section{Relativistic Cosmology}

In the application to  the  Cauchy-Kowalevski  theorem in   General Relativity, the condition \rf{Nashflow} provides  the  velocity propagation of the  surface  geometry  in terms of  a  quantity  which is independent  of  the metric  and of  coordinate choices.
  Since  $k_{ij}$ is a  rank-2  tensor, then it  also corresponds to a  spin-2 field,   defined by the   Fierz-Pauli dynamics. However, we  have seen  that  such  dynamics always   leads to  Einstein's  equations  when  the mass  $m$ in  \rf{FierzPauli} is  zero, so that  $k_{ij}$  must correspond  to a   massive spin-2 field defined by  the  Fierz-Pauli Lagrangian  defined in a  3-dimensional surface of  space-time,  obtained from \rf{FierzPauli}  with  $H_{ij}=k_{ij}(x,t)$,  $H_{i4}=0$ and  $H_{44}= h(x,t)$,  where  $h(x,t)$ is the mean  curvature of  the  surface
		\begin{equation}
{\cal L}^{mFP}\!\! =\!\!\frac{1}{4}\left [ h_{,i}h^{,j}\! -\! k_{jk;i}k^{jk;i} \! -\! 2k_{ij}{}^{;i}h^{,j}\!  +\! 2 k_{jk;i}k^{ik;j}\!-\! m^2(K^2\! -\! h^2) \right]\sqrt{-g}.   \label{eq:FierzPaulik}
\end{equation}
Here $K^2  =k^{ij}k_{ij}$  is the  Gaussian Curvature  of the surfaces and their  mean curvature of   are  defined by $h =g^{ij}k_{ij}$.  The indicated covariant derivatives  are calculated  with the  metric $g_{\mu\nu}$ of the space-time  and  its  surface- induced metric and  connection.

 The field equations for $k_{ij}$ are
\begin{equation}
(\Box^2 -m^2)\Psi_{ij}=0,  \hspace{4mm} \Psi_{ij}=k_{ij}-hg_{ij}, \label{eq:massivewave}
\end{equation}
where  $\Box^2 =g^{ij}\nabla_i\nabla_j$,  and $\Psi_{ij}$  may be called the \textit{deformation wave  function} so as   not to be   confused  with massive gravity.

Since  the  dynamics  of  $k_{ij}$ is independent of  the  metric, the  field  $k_{ij}$  does not enter in the geometric  side of  Einstein's  equations but it contributes  as  a  source  field through its  energy  tensor,  derived from the  potential energy  of  \rf{FierzPaulik}:
\begin{equation}
T^{mFP}_{ij} = \frac{m^2}{2}\left( k_i^m k_{mj}-h k_{ij}-\frac{1}{4}(K^2-h^2)g_{ij}\right). \label{eq:TmFPk}
\end{equation}
%

In  cosmology the  line  element of the   Friedmann, Lemaitre, Robertson,  Walker (FLRW) standard  model,  written in a  coordinate system  which is  co-moving  with the  Cauchy surfaces, can be  written as
\begin{equation}
ds^2 = -dt^2 + a(t)^2(dr^2 + f^2(r)r^2d\omega^2),
\label{eq:FLRW}
\end{equation}
where $f(r) = (sinhr, r, sinr)$,  corresponding respectively  to open, flat and  closed  universes (equivalently,  for $dr\rightarrow  \frac{dr}{1-\kappa r^2}$,  with  $\kappa= (-1, 0, 1)$).
Replacing the  FLRW metric components in  Einstein's equations and  considering only the perfect  fluid  as  a  source  we obtain the standard   Friedmann's equation for
\begin{equation}
\dot{a}^2 + \kappa =\frac{8\pi G}{3}\rho a^2.
\label{eq:Friedmann}
\end{equation}
Notice that the  standard derivation of  Friedmann's equation  does not include an  initial surface  and an  initial data  required by  the Cauchy-Kowalevski theorem. Any attempt to add such data at this  stage by using \rf{Christoffelflow} or \rf{Ricciflow}  would make the  propagation  velocity  entirely dependent  of  the  metric  and  consequently, it may induce errors.   In the following  we start anew,  regarding the    FLRW  cosmology, as  a  Cauchy problem with the use of  \rf{Nashflow}.

Assuming that the  inflation  occurred  isotropically,  the  Last  Inflationary Surface (LIS for  short),  just before  as  the universe entered the  FLRW period,   would be  a  3-sphere,  which is therefore taken  as the  initial surface  for  the relativistic  universe. An spherical initial surface naturally leads to  Legendre polynomials, which    form an orthogonal  and  complete  Fourier basis), independently  of  further boundary  conditions.

With  the presence of the   energy-momentum tensor  in the  right hand  side of  Einstein's  equations  we obtain
\begin{equation}
R_{\mu\nu} =-8pi G (T_{\mu\nu} -\frac{1}{2}T g_{\mu\nu} ) +T^m_{\mu\nu}-\frac{1}{2}T^m g_{\mu\nu}
\end{equation}
%
These  equations can  be  solved  for the FLRW metric in the same  way  the original  derivation of  Friedmann's  equation. The difference is  that here  we  have  a  Cauchy problem  with  initial value  for  $k_{ij}$  at  the LIS specifying that  $k_{ij}\rfloor_{LIS} \propto g_{ij}\rfloor_{LIS}$,  which   implies that the LIS is  a constant curvature  surface\cite{Eisenhart}.

After  the LIS,   the  observations  show that the  Cauchy surfaces  are  flat.  Therefore,  they  must  result from a   continuous deformation  from the LIS,   generating a sequence of   flat or  nearly flat surfaces.  This may  be  obtained with the supposition that
$k_{ij}= \alpha(t) g_{ij}$ (the proportionality factor is  a function of  time). More  specifically,  for the FLRW metric we may write 
\[
k_{ij} =\frac{b(t)}{a^2(t)}g_{ij}.
\]
where $b(t)=k_{11}$  remains an arbitrary function,    representing  the  radial  tension on the  surface.
With this  notation, the   energy tensor  resulting from  \rf{TmFPk} for the FLRW universe becomes
\begin{equation}
T^{mFP}_{ij} = \frac{m^2}{4} \frac{b^2}{a^2}g_{ij},
\label{eq:TmFPFkFried}
\end{equation}
It is more convenient to  define the  cosmological function $\Lambda(t)$ by
\begin{equation}
\frac{m^2}{4}\frac{b^2}{a^2}\stackrel{def}{=}\frac{\Lambda(t)}{3},  \label{eq:Lambda}
\end{equation}
so that after replacing  in  Einstein's  equations   we   obtain the  modified   Friedmann's  equation
\begin{equation}
\dot{a}^2 + \kappa  =\frac{8\pi G}{3}\rho a^2 - \frac{\Lambda(t)}{3}a^2.
\end{equation}
The  value of   $\Lambda (t)$  can now  be  determined by the  observed  acceleration of the universe,  which is  indicated by the observations  and that fits the $\Lambda$CDM paradigm  as  a  cosmological constant.
\[
\Lambda(t)\rfloor_{\mbox{observed}}\approx 10^{-47}Gev^4 \sim 10^{-29}g/cm^3.
\]
Replacing this  value  in  \rf{Lambda} and
 using the same  $\Lambda$CDM paradigm,
		we may  determine the  mass  $m$  from the estimates  for  dark matter.  For  example using the   Weakly Interacting Massive Particle (WIMP) dark matter model the value of  $m$ would  be   $m\approx  100$Gev.
Using this  mass  in \rf{Lambda}, the   component  $k_{11} \; (=b(t))$  may be determined.  		
Therefore  all  components of  the   Cauchy problem for the  present  relativistic phase of the universe are  determined.

\section{A Distant  Mirror}

The existence of an inflationary phase of the universe  was proposed to correct the predicted  age of the   FLRW universe  as  compared with the observed  age of  some  stars.
The  supposition  was that during the period of
$\approx 10^{-32}$ seconds, the  volume of the universe  increased  exponentially, something very  unlike  the  slow expansion observed  today. As we   have  seen, one such  rapid  growth of  volume  can be  explained by an  exponential function of the entropy.  In particular  for  $\sqrt{g}  =e^u$, we  obtain Ricci flow  deformation given by \rf{Ricciflow}.  Since this  is not  compatible with   Einstein's  equations used in the relativistic phase,  we  are in presence of  two  distinct  geometries,   with a proper junction condition between   them at the LIS.

It is  possible  that  the  inflationary phase  can  be  explained as  a Cauchy problem,
  by   establishing   an  initial surface,  say at the  Big Bang horizon, regarded as  a   compact and closed surface into which all  time-like geodesics  converge. Assuming that  the   expansion  of  the  inflationary  universe is  regulated by  the Ricci flow \rf{Ricciflow},  then by   a  similar   demonstration of the  Poincar\'e conjecture,   the   Big Bang horizon, deforms into the LIS spherical surface.
In that case,  the  LIS  would be  a  junction between  a geometry  defined by   $R_{ij}$ and  another  defined by Einstein's  equations and the  Nash flow  \rf{Nashflow},  that is by  $k_{ij}$.

The  Israel-Lanczos  junction  condition    is compatible  with both  thermal and  relativistic dynamics,   because it  compares  the  behavior  of  the  tangent and  normal  components of the  two geometries  when  passing from one  side of the LIS to the other \cite{Israel}. To understand this note that the  inflationary side of LIS, the Ricci  curvature of the  surface  is  defined only by  the  variations of the   tangent  vectors,  so that there is no change. However, in the relativistic side   the extrinsic  curvature depends on the  variation of the normal  vector  to  the  surface, which  varies  from one  side to  another. Therefore the junction  condition   at the  LIS requires that
	\begin{equation}
		R_{ij}\rfloor_{\mbox{LIS}} =k_{ij}\rfloor_{\mbox{LIS}}.
\end{equation}
This condition is  compatible  with  Einstein's  equations,   only
if  we impose that  the   surface has  a  $Z_2$ symmetry,  or more intuitively if  it  acts  as a mirror,  leading to the  expression  known as the Israel-Lanczos  condition  (For a detailed derivation see  \cite{MaiaAlcaniz})
\begin{equation}
k_{ij} =-8\pi G(T^{mFP}_{ij}-\frac{1}{3} T^{mFP}\;g_{ij}),
\label{eq:Israel}
\end{equation}
which  gives the extrinsic  curvature of  the LIS in terms of the   energy tensor of the then  existing  gravitational source,  the massive spin-2 field. Since the  inflation  happened before the structure formation of  ordinary matter  and  even  before the  radiation  condensation period of  the universe,  atoms  and gauge forces cannot  compose  that energy-momentum tensor  in  \rf{Israel}.
The only remaining  possibility is that  such  energy-momentum tensor results from  the existence of the  massive Fierz-Pauli spin-2 field  given by \rf{TmFPk}.
\vspace{2mm}\\	
The  existence of  such massive spin-2  field can be  explained by  today's observation of  the  acceleration of the universe  and the  lack of  mass  galaxies.

The existence  of  a   massive  spin-2  field  acting as  a  dark matter  content may  justify the additional gravitational pull,   stopping the  currently observed   acceleration, thus avoiding the  big rip scenario for end of the universe. Quite on the contrary,  the present universe  may  follow the arguments  similar  to an ``inverse Poincar\'e  conjecture'',   whereby  the  compact and  closed LIS  can  smoothly deform   into another  compact closed  surface (and   never into an open universe),  because  a  closed  compact space  cannot  be  smoothly deformed  into  an open volume. Clearly, this  requires  a relativistic generalization of  the proof of the Poincar\'e conjecture  using  the Nash flow  instead of  the Ricci flow.

In conclusion,  the  Cauchy-Kowalevski theorem  was  reviewed and applied to General Relativity and  cosmology,  using two  new tools: The ''geometrical analysis'' which  eliminates the necessity to use of analytic functions by the use of  the concept of  geometric flow. More importantly it leads us to  the use of  extrinsic curvature as a  massive  spin-2  field, required to complete  the set of  independent initial data required by the theorem.
In the beginning, the universe  was described as being  predominantly  thermal, promoting an  exponential growth of  the volume of the space-time with the  temperature, with the  Ricci-flow  geometry. This  was followed by today's  standard relativistic  universe,  which is compatible with  Nash flow geometry. The free  parameters  were  adjusted by the available observational data  within the  $\Lambda$CDM paradigm. The    additional massive spin-2  field   required by the Cauchy-Kowalevski theorem is  also regarded  as  a  possible  candidate to  dark matter,  while its  energy tensor  contributes to  dark energy.
The Last Inflationary Surface (LIS)  joins  the thermal and  relativistic  phases of  the universe,  acting as a  boundary  with a  mirror symmetry. The existence of  such symmetry  suggests that  the  present universe can be regarded  as  a reflection of  a past universe

\textbf{Acknowledgements:}

The  author  wishes  to  acknowledge professor  Stanley Deser for comments  and suggestions on a earlier  version of  this paper.

\end{document}